\documentclass[prd,preprint,tightenlines,floatfix,showpacs,preprintnumbers,nofootinbib,eqsecnum]{revtex4}
\usepackage{amsmath,amssymb,amsbsy}
 \usepackage[dvips,final]{graphicx}
 \usepackage{amsfonts}
 \usepackage{epsfig}
 \usepackage{bm}

\newcommand{\GeV}{\textrm{GeV}}
\newcommand{\MeV}{\textrm{MeV}}

\def\al{\relax\ifmmode\alpha\else{$\alpha${ }}\fi}
\def\alps{\relax\ifmmode\alpha_s\else{$\alpha_s${ }}\fi}
\def\as{\relax\ifmmode\alpha_s\else{$\alpha_s${ }}\fi}
\def\msbar{\relax\ifmmode\overline{\rm MS}\else{$\overline{\rm MS}${ }}\fi}
\newcommand{\ac}{\mathcal{A}}
\def\acal{\relax\ifmmode{\cal A}\else{${\cal A}${ }}\fi}

\begin{document}
\thispagestyle{empty} \preprint{\hbox{}} \vspace*{-10mm}

\title{Nucleon spin structure at low momentum transfers}

\author{Roman~S.~Pasechnik}
\email{roman.pasechnik@fysast.uu.se} \affiliation{High Energy
Physics, Department of Physics and Astronomy, Uppsala University Box
516, SE-75120 Uppsala, Sweden}

\author{Jacques Soffer}
\email{jacques.soffer@gmail.com}
\affiliation{Physics Department, Temple University\\
Barton Hall, 1900 N, 13th Street\\
Philadelphia, PA 19122-6082, USA}

\author{Oleg~V.~Teryaev}
\email{teryaev@theor.jinr.ru}
\affiliation{Bogoliubov Laboratory of Theoretical Physics, JINR,
Dubna 141980, Russia}

\date{\today}

\begin{abstract}

The generalized Gerasimov-Drell-Hearn (GDH) sum rule is known to be
very sensitive to QCD radiative and power corrections. We improve
the previously developed QCD-inspired model for the $Q^2$-dependence
of the GDH sum rule. We take into account higher order radiative and
higher twist power corrections extracted from precise Jefferson Lab
data on the lowest moment of the spin-dependent proton structure
function $\Gamma_1^{p}(Q^2)$ and on the Bjorken sum rule
$\Gamma_1^{p-n}(Q^2)$. By using the singularity-free analytic
perturbation theory we demonstrate that the matching point between
chiral-like positive-$Q^2$ expansion and QCD operator product
$1/Q^2$-expansion for the nucleon spin sum rules can be shifted down
to rather low $Q\simeq\Lambda_{QCD}$ leading to a good description
of recent proton, neutron, deuteron and Bjorken sum rule data at all
accessible $Q^2$.
\end{abstract}

\pacs{11.10.Hi, 11.55.Hx, 11.55.Fv, 12.38.Bx, 12.38.Cy}

\maketitle

\section{Introduction}

The problem of the nucleon spin structure and the peculiarities of
its underlying QCD description has attracted a lot
of attention over the recent years \cite{Anselmino:1994gn,Leader08}. In particular,
this is due to an
enormous progress in experimental studies of the spin sum rules at
low momentum transfer $Q^2$, from the very accurate
Jefferson Lab data on the lowest moment of the spin-dependent proton
structure function $\Gamma_1^p(Q^2)$ and on the Bjorken sum rule
$\Gamma_1^{p-n}(Q^2)$ in the range $0.05<Q^2<3\,{\rm GeV}^2$
\cite{JLab08data}. This data provided a good testing ground for combining both
the perturbative and non-perturbative QCD contributions.

Theoretical description of the nucleon spin structure functions
$g^{p,n}$ at large $Q^2$ relies on the Operator Product Expansion,
and at moderate $Q^2$ their
sensitivity to the radiative
 and higher twist power corrections becomes significant \cite{Larin}.
Due to such a sensitivity the
transition to the entirely non-perturbative $Q^2$ region is rather
cumbersome.
This transition was earlier addressed in
the QCD-motivated model \cite{ST} for the $Q^2$-dependence of
the generalized Gerasimov-Drell-Hearn (GDH) sum rule \cite{GDH}
making use of the relation to the Burkhardt-Cottingham sume rule
\cite{BC} for the structure function $g_2$, whose elastic
contribution is the main source of a strong $Q^2$-dependence, while
the contribution of the transverse structure function,
$g_T=g_1+g_2$, is smooth. The successful prediction of this model
was the distinct ``crossover'' point of the proton data for
$\Gamma_1^p(Q^2)$ at low $Q^2\sim 200\,-\,250\,\MeV^2$. Its
 subsequent modification \cite{ST1}, including radiative and power
QCD corrections, made the description far more accurate, which was
required by the increased accuracy of the data.

Now we enter a new level of increasing experimental accuracy, obtained in the
recently published proton JLab data \cite{JLab08data}. They lie
above the model inputs at $Q^2\gtrsim 1.5\,\GeV^2$ (while
displaying quite a similar shape) due to a noticeable sensitivity of
pQCD part of $\Gamma^{p,n}(Q^2)$ and to poorly known higher twist
contributions $\mu_{4,6,..}$, as well as the axial singlet charge
$a_0$.

Our present goal is to improve the model for the generalized GDH sum
rule for proton and neutron using the values of the power
corrections $\mu_{4,6,..}$ and singlet axial charge $a_0$,
systematically extracted from the JLab data \cite{PST08,PST09} and
by performing a similar program of the smooth interpolation between
large $Q^2$ and $Q^2=0$. As we will see we are able to achieve a
rather good description of the data at all $Q^2$ values.

The JLab data were obtained in the low $Q^2$ region and, therefore, a
special attention is needed to the QCD coupling in this domain.
While the $1/Q^2$ term in the OPE works at relatively high scales
$Q^2\gtrsim 1\,\GeV^2$, higher-twist (HT) power corrections
$1/Q^4,\,1/Q^6,$ etc., start to play a significant role at lower
scales, where the influence of the ghost singularities in the
coefficient functions within the standard perturbation theory (PT)
becomes more noticeable. It affects the results of extraction of the
higher twists from the precise experimental data leading to unstable
OPE series and huge error bars \cite{PST08}. It seems natural that
the weakening or elimination of the unphysical singularities of the
QCD coupling would allow shifting the perturbative QCD (pQCD)
frontier to a lower energy scale and to get more exact information
about the nonperturbative part of the process described by the
higher-twist series \cite{PST09}.

In this investigation, in order to avoid the influence of unphysical
singularities at $Q=\Lambda_{QCD}\sim 400\,\MeV$, we deal with the
ghost-free analytic perturbation theory (APT) \cite{apt96-7} (for a
review on APT concepts and algorithms, see also
Ref.~\cite{Sh-revs}), which was recently proven to be an intriguing
candidate for a quantitative description of light quarkonia spectra
within the Bethe-Salpeter approach \cite{BSAPT}, as well as in the
recent higher-twist analysis of the deep inelastic scattering data
on the $F_2$ structure function \cite{F2}. For completeness, we
compare our results obtained with conventional PT and APT couplings
and, finally, discuss the related uncertainties and stability
issues.

\section{Formalism}

\subsection{OPE regime $Q^2>\Lambda_{QCD}^2$}

To recall the basic ideas of the approach let us consider the lowest
moments of spin-dependent proton and neutron structure functions
$g^{p,n}_1$ defined as
\begin{eqnarray}\label{eq1}
\Gamma_1^{p,n}(Q^2)=\int^1_0dx\, g^{p,n}_1(x,Q^2)\,,
\end{eqnarray}
From now on, it is understood that the elastic contribution at $x=1$ is
excluded from the moments, since it is the
``inelastic'' contribution which can be matched with
GDH sum rule.

At large $Q^2$ the moments $\Gamma_1^{p,n}(Q^2)$ are given by the
OPE series in powers of $1/Q^2$
with the expansion coefficients
(see, e.g., Ref.~\cite{kataev}). In the limit $Q^2\gg M^2$ the
moments are dominated by the leading twist contribution,
$\mu_2^{p,n}(Q^2)$, which can be decomposed into flavor singlet and
nonsinglet contributions:
\begin{eqnarray}
\Gamma^{p,n}_{1}(Q^2)=\frac{1}{12}\left[\biggl(\pm
a_3+\frac13a_8\biggr)E_{NS}(Q^2)+\frac43 a_0\,E_{S}(Q^2)\right]+
\sum_{i=2}^{\infty}\frac{\mu^{p,n}_{2i}(Q^2)}{Q^{2i-2}},\label{PT-Gam}
\end{eqnarray}
where $E_{S}$ and $E_{NS}$ are the singlet and nonsinglet Wilson
coefficients, respectively, calculated as series in powers of $\as$
\cite{Larin:1997qq}. These coefficient functions for $n_f=3$ active
flavors in the \msbar scheme are
\begin{eqnarray} \label{nonsinglet}
E_{NS}(Q^2)&=&1-\frac{\alpha_s}{\pi}-3.558\left(\frac{\alps}{\pi}\right)^2
-20.215\left(\frac{\alps}{\pi}\right)^3-O(\alps^4)\,, \\
E_{S}(Q^2)&=&1-\frac{\alpha_s}{\pi}-1.096\left(\frac{\alps}{\pi}\right)^2-O(\alps^3)\,.
\label{singlet}
\end{eqnarray}
The triplet and octet axial charges $a_3\equiv g_A=1.267\pm0.004$
\cite{pdg08} and $a_8=0.585\pm0.025$ \cite{Goto:1999by},
respectively, are extracted from weak decay matrix elements.
As for the singlet axial
charge $a_0$, it is convenient to work with its renormalization group (RG) invariant
definition in the \msbar scheme $a_0=a_0(Q^2=\infty)$, in which all
the $Q^2$ dependence is factorized into the definition of the Wilson
coefficient $E_S(Q^2)$. For detailed discussion of the higher-loop
stability of the coefficient functions and prescriptions used in
actual calculations, see Ref.~\cite{PST09}.

We address both proton and neutron spin sum rules (SSRs), and the
singlet and octet contributions are canceled out in their difference
$\Gamma_1^p-\Gamma_1^n$ resulting in the Bjorken sum rule \cite{Bj66}
\begin{eqnarray}
\Gamma^{p-n}_{1}(Q^2)=\frac{g_A}{6}E_{NS}(Q^2)+
\sum_{i=2}^{\infty}\frac{\mu^{p-n}_{2i}(Q^2)}{Q^{2i-2}}.\label{BSR}
\end{eqnarray}

The unphysical singularities at
$Q\sim \Lambda_{QCD}$ in the PT series for the coefficient functions
$E_{S}(Q^2)$ (\ref{singlet}) and $E_{NS}(Q^2)$ (\ref{nonsinglet})
strongly affect the analysis of the spin sum rules at low $Q^2$ \cite{PST09}.
Their influence becomes essential at $Q<1\,\GeV$ where the HT terms
start to play an important role. The ``soft-frozen'' $\alpha_s$
models are free of such a problem, thus providing a more reliable
tool of investigating the behavior of the spin sum rules in the
low-energy domain.

The moments of the structure functions are analytic functions in the
complex $Q^2$ plane with a cut along the negative real axis, as
demonstrated in Refs.~\cite{W78,Ashok_suri}. On the other hand, the
standard PT approach does not support these analytic properties. The
APT method \cite{apt96-7} gives the possibility of combining the RG
resummation with correct analytic properties of the QCD corrections.
The consequence of requiring these properties to hold in the DIS
description was studied previously in Refs.~\cite{APT-GLS,MSS}.

Let us recall that the expression for $\Gamma_1^{p,n}(Q^2)$ in the
framework of the analytic approach is completely similar to the one in
the standard PT (\ref{PT-Gam}):
\begin{eqnarray}
\Gamma^{p,n}_{1,APT}(Q^2)=\frac{1}{12}\left[\biggl(\pm
a_3+\frac13a_8\biggr)E^{APT}_{NS}(Q^2)+\frac43
a^{inv}_0\,E^{APT}_{S}(Q^2)\right]+
\sum_{i=2}^{\infty}\frac{\mu^{APT}_{2i;\,p,n}(Q^2)}{Q^{2i-2}} \, .
\label{APT-Gam}
\end{eqnarray}
The corresponding NNLO APT modification of the singlet and
nonsinglet coefficient functions is
\begin{eqnarray}\nonumber
E^{APT}_{NS}(Q^2)&=&1-0.318\,{\cal A}^{(3)}_1(Q^2)-0.361\,{\cal
A}^{(3)}_2(Q^2)-\,...\,, \\
E^{APT}_{S}(Q^2)&=&1-0.318\,{\cal A}^{(3)}_1(Q^2)-0.111\,{\cal
A}^{(3)}_2(Q^2)-\,...\, , \label{E-APT}
\end{eqnarray}
where ${\cal A}^{(3)}_k$ is the analyticized $k$-th power of three-loop
PT coupling in the Euclidean domain and defined as
 \begin{eqnarray}
  \label{Akn}
  \ac^{(n)}_k(Q^2)=\frac{1}{\pi} \int^{+\infty}_0
  \frac{\mathrm{Im}([\as^{(n)}(-\sigma,n_f)]^k)\,d\sigma}{\sigma+Q^2},\qquad
  n=3\,.
 \end{eqnarray}
In the one-loop case, the APT Euclidean functions are simple enough
\cite{apt96-7}:
\begin{eqnarray}
 &&\acal_1^{(1)}(Q^2)=\frac{1}{\beta_0}\left[\frac{1}{L}+
 \frac{\Lambda^2}{\Lambda^2-Q^2}\right]\,,\quad
 L=\ln\left(\frac{Q^2}{\Lambda^2}\right),\label{AE1-2}\\
 &&\acal_2^{(1)}(l)=\frac{1}{\beta_0^2}\left[\frac{1}{L^2}-
 \frac{Q^2\,\Lambda^2}{(Q^2-\Lambda^2)^2}\right],\;
 \acal_{k+1}^{(1)}=-\,\frac{1}{k\,\beta_0}
 \frac{d \acal_k^{(1)}} {d L}.\nonumber
\end{eqnarray}
Analogous two- and three-loop level
expressions are more involved. However, according to the ``effective
log'' approach \cite{SolSh99} in the region $\,Q<5\,\,\GeV$ one may
use simple one-loop expressions (\ref{AE1-2}) with the {\it
effective logarithm} $L^*\,$:
\begin{eqnarray} \label{model}                       
 \acal_{1,2,3}^{(3)}(L)\to\acal_{1,2,3}^{mod}=\,\acal_{1,2,3}^{(1)}(L^*)
 \,,\quad L^*\simeq 2\,\ln(Q/\Lambda^{(1)}_{\mathrm{eff}}),\quad
 \Lambda^{(1)}_{\mathrm{eff}}\simeq
 0.50\,\Lambda^{(3)}.
\end{eqnarray}
Thus, instead of the exact three-loop expressions for the APT
functions, in Eq.~(\ref{E-APT}) one can use the one-loop expressions
(\ref{AE1-2}) with the effective $\Lambda$ parameter
$\Lambda_{mod}=\Lambda^{(1)}_{\mathrm{eff}}\,$ whose value is given
by the last relation (\ref{model}). This model was successfully
applied for higher-twist analysis of low-energy JLab data in
Refs.~\cite{PST08,PST09}, and also in the $\Upsilon$ decay analysis
in Ref.~\cite{ShZ05}. Note  also that the APT
couplings are stable with respect to different loop orders at
low-energy scales $Q^2\lesssim 1\,\GeV^2$ \cite{Sh-revs}, contrary to
the standard PT approach.

The APT functions $\ac_k$ contain the $(Q^2)^{-k}\,$ power
contributions, which effectively change the values of the $\mu$-terms,
when going from the PT to the APT framework. In particular, by subtracting an
extra $(Q^2)^{-1}$ term induced by the APT series for the Bjorken
sum rule
\begin{eqnarray*}
&&\Gamma^{p-n}_{1,APT}(Q^2)\simeq\frac{g_A}{6}+
f\biggl(\frac{1}{\ln(Q^2/{\Lambda^{(1)}_{\mathrm{eff}}}^2)}\biggr)+
\varkappa\frac{{\Lambda^{(1)}_{\mathrm{eff}}}^2}{Q^2}+ {\cal
O}\left( \frac{1}{Q^4} \right)
\end{eqnarray*}
where $\varkappa\simeq0.43$ and
$\Lambda^{(1)}_{\mathrm{eff}}\sim0.18\,\GeV$ is the effective
one-loop $\Lambda_{QCD}$ parameter, we get the relation between
$\mu_{4,APT}^{p-n}$ coming into the APT expression (\ref{APT-Gam})
and the conventional $\mu^{p-n}_{4}$ from Eq.~(\ref{PT-Gam}):
\begin{eqnarray}                                
 \label{mu4_APT}
 \frac{\mu^{p-n}_4(1\,\GeV^2)}{M^2}\simeq\frac{\mu_{4,APT}^{p-n}+
 \varkappa{\Lambda^{(1)}_{\mathrm{eff}}}^2}{M^2}\,.
\end{eqnarray}

Along with the conventional PT scheme, we will also apply the APT
approach based on Eqs.~(\ref{APT-Gam}) and (\ref{E-APT}) to
construct the improved model for smooth continuation of perturbative
expressions for $\Gamma_1^{p,n}(Q^2)$ and its non-singlet
combination $\Gamma_1^{p-n}(Q^2)$ down to the non-perturbative
region $Q^2\to 0$.

\subsection{``Chiral'' regime $Q^2\lesssim\Lambda_{QCD}^2$}

For the purpose of a smooth continuation of $\Gamma_1^{p,n}(Q^2)$ to
the non-perturbative region $0\leq Q^2\lesssim \Lambda_{QCD}^2$
\cite{ST}, we consider firstly the $Q^2$-evolution of the integral
\begin{eqnarray}\label{I1}
I_1(Q^2)\equiv\frac{2M^2}{Q^2}\Gamma_1(Q^2)=\frac{2M^2}{Q^2}\int^1_0dx\,
g_1(x,Q^2)\,,
\end{eqnarray}
which is equivalent to the integral over all energies of the
spin-dependent photon-nucleon cross-section, whose value at $Q^2=0$
is defined by the GDH sum rule \cite{GDH}
\begin{eqnarray}\label{I10}
I_1(0)=-\frac{\mu_A^2}{4}\,,
\end{eqnarray}
where $\mu_A$ is the nucleon anomalous magnetic moment. Then, the
function $I_1(Q^2)$ can be written as a difference
\begin{eqnarray}\label{dec}
I_1(Q^2)=I_T(Q^2)-I_2(Q^2),
\end{eqnarray}
where
\begin{eqnarray}\label{IT2}
I_T(Q^2)=\frac{2M^2}{Q^2}\int^1_0dx\, g_T(x,Q^2),\qquad
I_2(Q^2)=\frac{2M^2}{Q^2}\int^1_0dx\, g_2(x,Q^2)\,.
\end{eqnarray}

The well-known Burkhardt-Cottingham (BC) sum rule \cite{BC} provides
us with an exact expression for $I_2(Q^2)$, in terms of familiar
electric $G_E$ and magnetic $G_M$ Sachs form factors as
\begin{eqnarray}\label{I2}
I_2(Q^2)=\frac14\mu G_M(Q^2)\frac{\mu
G_M(Q^2)-G_E(Q^2)}{1+Q^2/4M^2}\,,
\end{eqnarray}
where $\mu$ is the nucleon magnetic moment. As a consequence of the
strong $Q^2$ behavior of the r.h.s. of Eq.~(\ref{I2}), we get for
large $Q^2$
\begin{eqnarray}\label{I2Q2}
\int_0^1g_2(x,Q^2)dx\big|_{Q^2\to\infty}=0\,,
\end{eqnarray}
so $I_2$ is much smaller than $I_1$ for large $Q^2$. Now from the BC sum
rule (\ref{I2}), it follows that
\begin{eqnarray}\label{I20}
I_2(0)=\frac{\mu_A^2+\mu_A e}{4}
\end{eqnarray}
where $e$ is the nucleon charge. Then the GDH value (\ref{I10}) is
reproduced with
\begin{eqnarray}\label{IT0}
I_T(0)=\frac{\mu_A e}{4}\,.
\end{eqnarray}
To summarize, from the above equalities (\ref{I2}), (\ref{I2Q2}) and
(\ref{IT0}), we can conclude that the BC and GDH sum rules together,
lead to positivity of $I_T(Q^2)$ for all $Q^2$ in the proton case
and a vanishing difference between $I_T(Q^2)$ and $I_1(Q^2)$ for large
$Q^2$. Thus, $I_T^p(Q^2)$ is a smooth and monotonous function, and it
is possible to obtain its smooth interpolation between large $Q^2$
and $Q^2=0$ \cite{ST}.

\section{Improved model for smooth interpolation of $I_T(Q^2)$}

To improve the agreement between the model predictions and the
experimental data, we consider the general asymptotic expression
\begin{eqnarray}
I^{p,n}_{1,pert}(Q^2)=\frac{2M^2}{Q^2}\left[\frac{1}{12}\biggl(\pm
a_3+\frac13a_8\biggr)E_{NS}(Q^2)+\frac19 a_0\,E_{S}(Q^2)+
\sum_{i=2}^{\infty}\frac{\mu^{p,n}_{2i}(Q^2)}{Q^{2i-2}}\right],\label{I1pert}
\end{eqnarray}
where the nonsinglet $E_{NS}$ and singlet $E_S$ coefficient functions
are defined in Eqs.~(\ref{nonsinglet}) and (\ref{singlet}),
respectively. Then, the perturbative expression for $I_T$, defined
above the matching point $Q_0^2$, is
\begin{eqnarray}\label{ITpert}
I_{T,pert}(Q^2)=\Theta(Q^2-Q_0^2)\,[I_{1,pert}(Q^2)+I_2(Q^2)]\,,
\end{eqnarray}
where $I_{1,pert}(Q^2)$ is calculated from Eq.~(\ref{I1pert}), while
$I_2(Q^2)$ in known from the BC sum rule (\ref{I2}). The smooth
interpolation to the GDH value at $Q^2=0$ (\ref{IT0}) is difficult
and cannot be performed analytically. Following the procedure
developed in Ref.~\cite{ST}, we instead make use of the smooth
extrapolation of the perturbative expression (\ref{ITpert}), to the
nonperturbative domain $Q^2<Q_0^2$ defining the polynomial in
positive powers of $Q^2$ as
\begin{eqnarray}\label{ITnpert}
I_{T,nonpert}(Q^2)=\Theta(Q_0^2-Q^2)\,\sum_{n=0}^N\,\frac{1}{n!}
\frac{\partial^n\,I_{T,pert}}{\partial(Q^2)^n}\Big|_{Q=Q_0}(Q^2-Q_0^2)^n\,,
\end{eqnarray}
where $N$ is the number of derivatives, which is a free parameter of
the model, together with the matching point $Q=Q_0$, which have to
be chosen to satisfy
\begin{eqnarray}\label{ITconst}
I_{T,nonpert}(0)=\frac{\mu_A e}{4}\,.
\end{eqnarray}
In practice, the easiest way to solve the problem is to fix the
number of derivatives $N$ and then to vary the $Q_0$ value until the
relation (\ref{ITconst}) is satisfied. It is interesting to note
that taking $N=1$ does not allow for such a solution.

Such a procedure can be considered as a matching of the
``twist-like'' expansion in negative powers of $Q^2$ and the
``chiral-like'' expansion in positive powers of $Q^2$ \cite{ST},
which is similar to the matching of the expansions in direct and
inverse coupling constants.

Once we have obtained the parameters $N$ and $Q_0^2$, then the
all-$Q^2$ expressions for the moments $\Gamma_1^{p,n}$ can be
restored from $I_T(Q^2)$ defined by Eqs.~(\ref{I1pert}) --
(\ref{ITnpert}), by using Eqs.~(\ref{I1}) and (\ref{dec}).  As we
will see below, this can be done within both the standard PT and
singularity-free APT in the same way, leading to rather similar
curves for $Q^2$-evolution, except that in the APT case the matching
point $Q_0^2$ playing a role of the ``pQCD frontier'' in this
interpolation scheme is noticeably shifted down to lower $Q^2$
scales (see the next Section).

\section{Higher-twist analysis}

A detailed higher twist analysis of the recent Jefferson Lab data
\cite{JLab08data} on the lowest moments of the spin-dependent proton
and neutron structure functions $\Gamma_1^{p,n} (Q^2)$ and
$\Gamma_1^{p-n}(Q^2)$ in the range $0.05<Q^2<3\,{\rm GeV}^2$ was
performed in Refs.~\cite{PST08,PST09}. In particular, including only
three terms of the OPE expansion $\mu_{4,6,8}$ in
Eq.~(\ref{PT-Gam}), a satisfactory description of the data has been
achieved down to $Q^2\simeq0.17\,\GeV^2$ in conventional PT and down
to $Q^2\simeq0.10\,\GeV^2$ in the APT.

The lower $Q^2$ involved, the higher twist contribution is needed to
describe the data. As was shown in Ref.~\cite{PST09}, there is some
sensitivity of fitted values of $\mu_4$ to the minimal scale $Q_{min}$
variations; namely, it increases in magnitude when one incorporates
into the fit the data points at lower energies. This property of the
fit was treated as the slow (logarithmic) evolution $\mu_4(Q^2)$
(and $a_0(Q^2)$ in the singlet case) with $Q^2$ which becomes more
noticeable for broader fitting ranges in $Q^2$, as discussed above.
Indeed, fit results for $\mu_4$ with taking into account the RG
evolution with $Q=1\,\GeV$, as a normalization point become more
stable with respect to $Q_{min}$ variations.

However, there is still a problem how to treat the evolution of
higher-twist terms $\mu_{6,8,..}(Q^2)$ which again may turn out to
be important when one goes to lower $Q^2$, since the fit becomes
more sensitive to very small variations of $\mu_{6,8,..}$ with
$Q^2$. Since the evolution of the higher twists $\mu_{6,8}$ is still
theoretically unknown, they can be taken as free parameters
\cite{Deur-fits}. This procedure leads to rather small
$\chi^2_{D.o.f}\sim 0.1$ since more free parameters come into a fit
at lower $Q^2$. On the other hand, by including $\mu_{6,8,..}$ into
the fit, one observes only a small change in
$\mu_4(1\,\GeV^2)$~\cite{PST09}, which demonstrates its stability
down to lower $Q^2$. Taking this into account, in order to reduce
the number of free parameters, in the current work we apply another
fitting procedure and determine, first, $\mu_4$ and $a_0$ at higher
scale $Q=1\,\GeV$. Then we extract $\mu_6$ applying the known QCD
evolution for $\mu_4(Q^2)$ and $a_0(Q^2)$ and fixing them at
$1\,\GeV$ from the previous fit. The number of free parameters does
not grow in this case. The fitting domain is restricted from below
by $Q_{min}$ defined by the condition $\chi^2/D.o.f\leq1$. The
corresponding results are listed in Table~\ref{tab:Bjtot}. Due to
unknown evolution of $\mu_6(Q^2)$, which tends to be quite
noticeable at lower $Q^2$, we do not go below $Q_{min}$ and do not
take into account $\mu_8$-term here.

The advantage of the APT analysis is the infrared and higher loop
stability of the radiative corrections, as well as the stability
w.r.t. $\Lambda_{QCD}$ variations, leading to the stability and
convergence of the higher twist series extracted from the data.
Indeed, as we see from Table~\ref{tab:Bjtot}, in the APT case the
applicability of the perturbative expansion (\ref{APT-Gam}) is
somewhat shifted down to lower $Q^2$, due to the absence of Landau
singularities (see also Ref.~\cite{PST09} and references therein).

\section{All-$Q^2$ spin sum rules}

In the perturbative expression (\ref{I1pert}) we take into account
the two-loop perturbative correction in the singlet $E_S$ and
non-singlet $E_{NS}$ coefficient functions, as well as the twist-4,6
contributions discussed in the previous section. To explore the
infrared sensitivity of the model of the smooth continuation to
$Q=0$, we used two different sets of higher twist terms (with
$\mu_4$ and $\mu_{4,6}$, respectively) and the corresponding singlet
axial charge extracted from the data above a certain minimal scale
$Q_{min}^2$.
\begin{figure}[!h]
\begin{minipage}{0.49\textwidth}
 \centerline{\includegraphics[width=1.0\textwidth]{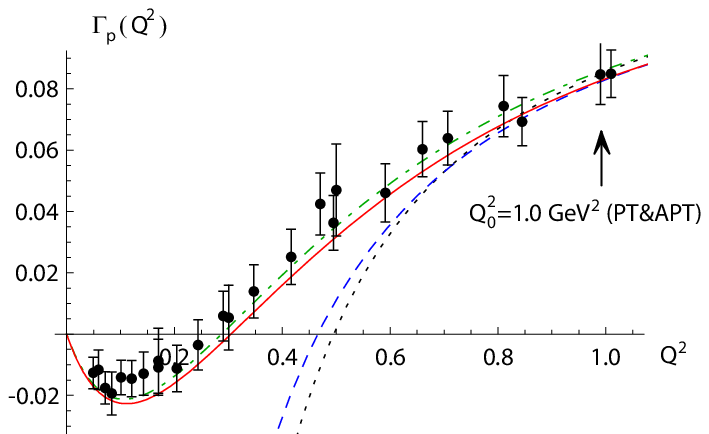}}
\end{minipage}
\begin{minipage}{0.49\textwidth}
 \centerline{\includegraphics[width=1.0\textwidth]{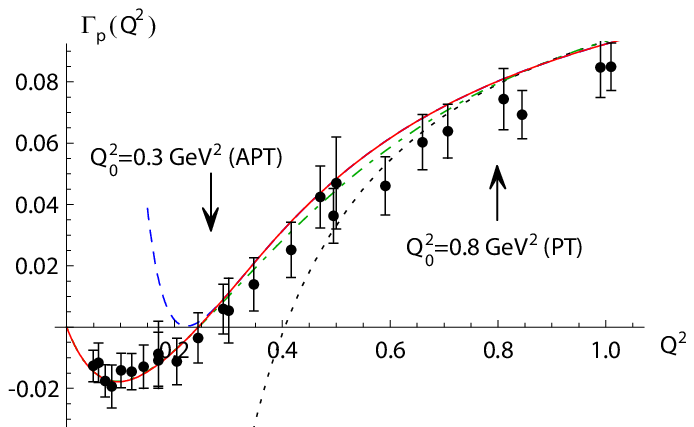}}
\end{minipage}
   \caption{\label{fig:proton}
   \small \em
   Proton spin sum rule function $\Gamma_1^p(Q^2)$ with respect to
   the combined set of JLab and SLAC data.
   Results are shown with an account of the twist-4 term (left panel) and the twist-4,6 terms
   (right panel). Corresponding perturbative parts are calculated
   in the framework of conventional PT (dotted lines) and APT
   (dashed lines). All-$Q^2$ model function obtained by the smooth
   interpolation of $I_T^p(Q^2)$ is also presented in PT
   (dash-dotted lines) and APT (solid lines).}
\end{figure}

\begin{figure}[!h]
\begin{minipage}{0.49\textwidth}
 \centerline{\includegraphics[width=1.0\textwidth]{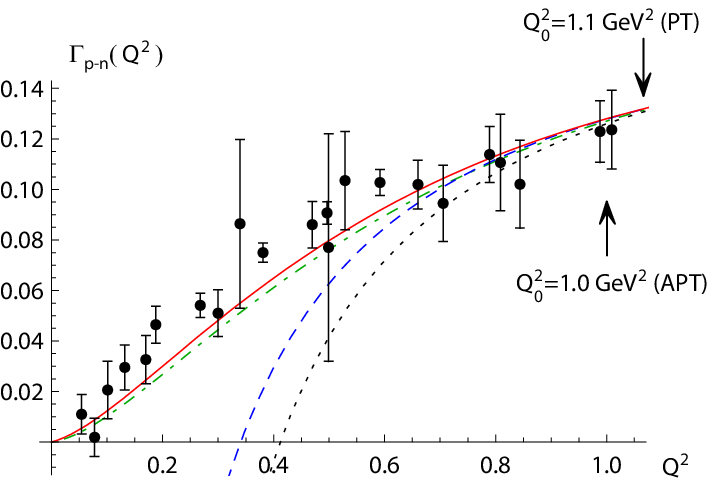}}
\end{minipage}
\begin{minipage}{0.49\textwidth}
 \centerline{\includegraphics[width=1.0\textwidth]{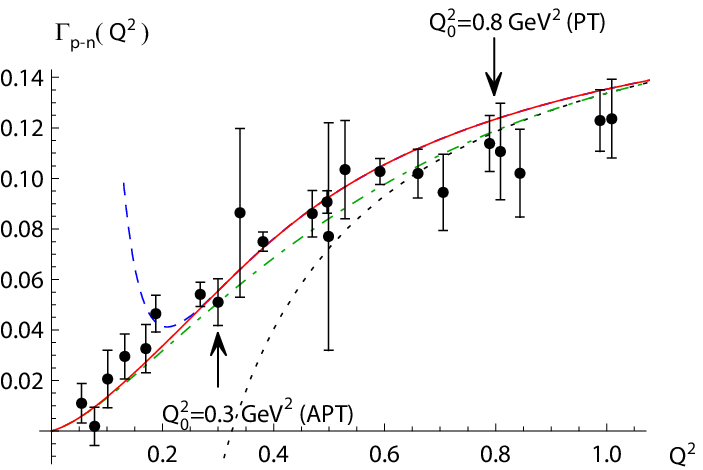}}
\end{minipage}
   \caption{\label{fig:BSR}
   \small \em  Bjorken sum rule function $\Gamma_1^{p-n}(Q^2)$ with respect to
   the combined set of JLab and SLAC data.
   The meaning of curves here is the
   same as in Fig.~\ref{fig:proton}.}
\end{figure}

In Fig.~\ref{fig:proton} we present the proton spin sum rule
function $\Gamma_1^p(Q^2)$ obtained by the smooth interpolation of
the perturbative part $I_{T,perp}^p(Q^2)$ to the non-perturbative
region $Q^2\to0$ as described in Section III. Calculations taking
into account one twist-4 term (left panel) and two twist-4,6 terms
(right panel) listed above are performed within the conventional PT
and APT. In Fig.~\ref{fig:BSR} we show the Bjorken sum rule function
$\Gamma_1^{p-n}(Q^2)$ calculated at any $Q^2$ in the similar way as
$\Gamma_1^p(Q^2)$.
\begin{table}[!h]
\caption{\small\sf Combined fit results of JLab and SLAC data on the
Bjorken SR and proton SSR for the singlet axial charge $a_0$, and
the higher-twist terms $\mu_4$ and $\mu_6$ defined at the
normalization point $Q^2=1\,\GeV^2$ in the APT and the standard PT
approaches, along with the matching value $Q_0^2$. Corresponding
curves for $\Gamma_1^p(Q^2)$ and $\Gamma_1^{p-n}(Q^2)$ are shown in
Figs.~\ref{fig:proton} and \ref{fig:BSR}, respectively. Typical
values of $\chi^2/D.o.f$ are close to unity.}
\begin{center}\label{tab:Bjtot}
\begin{tabular}{|c|c|c|c|c|c|c|} \hline
 $\quad$Method$\quad$& $\quad$Target$\quad$& $Q_{min}^2,\,\GeV^2$ & $\; a_0\;$ &
 $\;\mu_4/M^2\;$ & $\quad\mu_6/M^4\quad$ & $\quad Q_0^2,\,\GeV^2\quad$   \\ \hline\hline
                     & p-n    & 1.0   &   --    & $-0.060(3)$  &     0     & 1.1(2) \\
      NLO PT         &        & 0.3   &   --    & $-0.060   $  &  0.010(2) & 0.8(2) \\ \cline{2-7}
                     & proton & 1.0   & 0.34(3) & $-0.056(3)$  &     0     & 1.0(1)  \\
                     &        & 0.5   &  0.34   & $-0.056   $  &  0.010(2) & 0.8(2) \\ \hline\hline
                     & p-n    & 1.0   &   --    & $-0.058(3)$  &     0     & 1.0(1) \\
      NLO APT        &        & 0.2   &   --    & $-0.058   $  &  0.010(1) & 0.3(1) \\ \cline{2-7}
                     & proton & 1.0   & 0.37(2) & $-0.063(2)$  &     0     & 1.0(1) \\
                     &        & 0.3   &  0.37   & $-0.063   $  &  0.011(1) & 0.3(1) \\ \hline\hline
\end{tabular} \end{center}
\end{table}

The all-$Q^2$ model functions $\Gamma_1^p(Q^2)$ and
$\Gamma_1^{p-n}(Q^2)$ in both versions of the perturbation theory
(dash-dotted and solid lines) are rather close to each other
demonstrating the agreement between the singularity-free APT
analysis at lower $Q^2$ and the usual PT one at relatively higher
$Q^2$. Also, as one can see from the comparison of the left and
right panels the results of the interpolation do not strongly depend
on the number of higher twists included and, hence, on the border
$Q_0^2$ between perturbative and non-perturbative regimes. This
exhibits a sort of duality between them implying that the
experimental data in the wide intermediate region
$\Lambda^2_{QCD}\lesssim Q^2\sim1\,\GeV^2$ can be described equally
well either by OPE $1/Q^2$-series or by ``chiral-like''
$Q^2$-series.

We studied the sensitivity of above results w.r.t. variations of the
number of derivatives $N$ in Eq.~(\ref{ITnpert}) being the number of
positive $\sim Q^{2i}$ power terms. As mentioned above, at lower
$Q^2$ we need more higher $1/Q^2$-power twist terms. In the same
way, going up from very low $Q^2$ we observe analogously that to
describe the data at higher $Q^2$ we need more $\sim Q^{2i}$ power
terms, i.e. a higher value $N$.

The minimal number of derivatives $N_{min}$, which is necessary to
perform the smooth extrapolation according to Eq.~(\ref{ITnpert}) in
the conventional PT case and with one $\mu_4$ term only, is
$N_{min}=4$. Corresponding matching value between perturbative and
non-perturbative domains in this case is found to be
$Q_0^2=1.0\pm0.1\,\GeV^2$ for the proton SSR and $Q_0^2=1.1\pm
0.2\,\GeV^2$ for the Bjorken SR (see Table~\ref{tab:Bjtot}).
However, if one increases the number of $Q^2$-power term up to
$N=6$, the applicability of the ``chiral-like'' expansion raises up
to $Q_0^2\simeq1.4\,\GeV^2$ for the proton SSR and
$Q_0^2\simeq1.5\,\GeV^2$ for the Bjorken SR. Similar observation was
made earlier in Ref.~\cite{ST}.

In the framework of APT the minimal number of derivatives
$N_{min}=3$ is even smaller than in the conventional PT. In this
case, if only one $\mu_4$ term is included then the matching value
turned out to be same as in PT: $Q_0^2\simeq1.0\,\GeV^2$ for both
the proton SSR and $p-n$ demonstrating the similarity of the APT and
PT predictions at $Q\gtrsim 1\,\GeV$.

However, analysis at lower $Q^2$ including two $\mu_{4,6}$ terms
leads to quite different $Q_0$ values for PT and APT. In this case,
for the proton SSR and Bjorken SR we have $Q_0^2\simeq
0.8\pm0.2\,\GeV^2$ (PT, $N_{min}=3$) and
$Q_0^2\simeq0.32\pm0.1\,\GeV^2$ (APT, $N_{min}=2$). Such a shift of
the border between perturbative and non-perturbative domains in the
APT is a direct consequence of the disappearance of the unphysical
singularities in the radiative corrections, and confirms the similar
conclusion made in Refs.~\cite{PST08,PST09}.
\begin{figure}[!h]
\begin{minipage}{0.49\textwidth}
 \centerline{\includegraphics[width=1.0\textwidth]{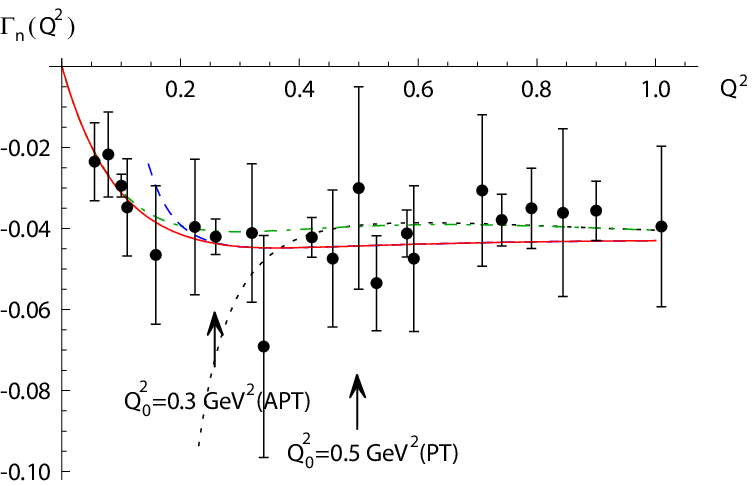}}
\end{minipage}
\begin{minipage}{0.49\textwidth}
 \centerline{\includegraphics[width=1.0\textwidth]{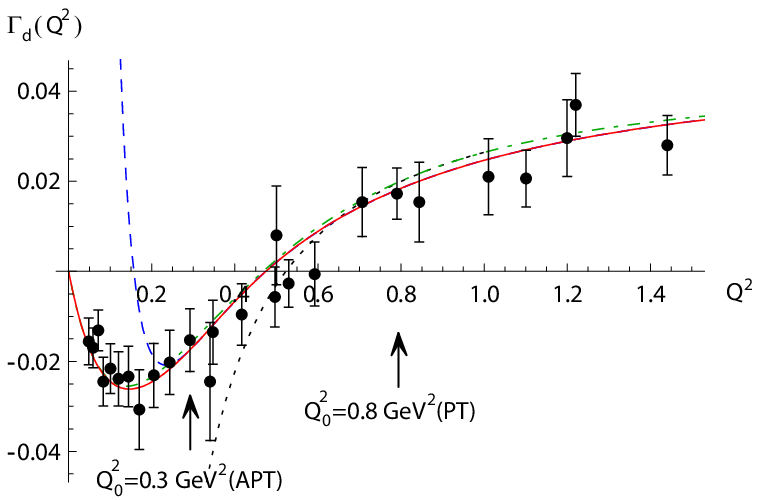}}
\end{minipage}
   \caption{\label{fig:nd}
   \small \em  Neutron (left) and deuteron (right) spin sum rule functions,
   $\Gamma_1^n(Q^2)$ and $\Gamma_1^d(Q^2)$, with respect to
   the combined set of JLab and SLAC data. Results are shown with
   an account of twist-4,6 terms. The meaning of curves here is the
   same as in Fig.~\ref{fig:proton}.}
\end{figure}

Finally, in Fig.~\ref{fig:nd} we show the neutron spin sum rule
function $\Gamma_1^n(Q^2)$, which is simply obtained from the
difference $\Gamma_1^p(Q^2)-\Gamma_1^{p-n}(Q^2)$, and the deuteron
spin sum rule $\Gamma_1^d(Q^2)$. We also present its perturbative PT
and APT parts together with less precise data. Both versions of the
perturbation theory predict monotonous curves for
$\Gamma_1^{n,d}(Q^2)$ at any $Q^2$. Comparison between them and the
results of Ref.~\cite{PST09} demonstrates the consistence of the
direct fits to the data and the predictions of the generalized GDH
sum rule.

\section{Conclusion}

In the current paper we have considered the all-$Q^2$ model for the
generalized GDH sum rule, constructed by the smooth interpolation of
$I_T(Q^2)$ between large $Q^2$ and $Q^2=0$, in the framework of both
the conventional PT and the ghost-free Analytic Perturbation Theory.
We used the values of the power corrections $\mu_{4,6,..}$ and
singlet axial charge $a_0$, systematically extracted from the
precise JLab data. We achieve a rather good description of the
proton data on $\Gamma_1^p(Q^2)$ at any $Q^2$ values. We also
present an improved description of the neutron data, as well as the
Bjorken sum rule data at all experimentally accessed $Q^2$.

The results of the smooth interpolation $\Gamma_1^p(Q^2)$ and
$\Gamma_1^{p-n}(Q^2)$ do not strongly depend on the number of
higher-twist terms, and on the border $Q_0^2$ between perturbative
and non-perturbative regimes. This exhibits a sort of duality
between them implying that the experimental data in the wide
intermediate region $\Lambda_{QCD}^2\lesssim Q^2\sim1\,\GeV^2$ can
be described equally well either by OPE $1/Q^2$-series or by
non-perturbative ``chiral-like'' $Q^2$-series. Within the analytic
PT the ``pQCD frontier'' being the matching value between $Q^2$- and
$1/Q^2$-power series naturally decreases from $1.0\,\GeV^2$ with
single $\mu_4$ down to $0.3\,\GeV^2$ with extra $\mu_6$-term
included, which is significantly lower than the corresponding value
in conventional PT $Q_0^2\simeq 0.8\,\GeV^2$. Such a shift of the
border between perturbative and non-perturbative domains in the APT
is a direct consequence of the disappearance of the unphysical
singularities in the radiative corrections.

\section*{ACKNOWLEDGMENTS}

We are thankful to A.P.~Bakulev, J.P.~Chen, G.~Dodge,
S.B.~Gerasimov, G.~Ingelman, A.L.~Kataev, O.V.~Selyugin,
A.V.~Sidorov, D.B.~Stamenov, and N.G.~Stefanis for valuable
discussions. This work was partially supported by the Carl Trygger
Foundation and by RFBR Grants No. 09-02-00732, 09-02-01149.

\end{document}